\documentclass[reprint, amsmath,amssymb,aps]{revtex4-2}

\usepackage{graphicx}
\usepackage{dcolumn}
\usepackage{bm}
\usepackage{amsmath}
\usepackage[usenames,dvipsnames]{xcolor}
\usepackage{mathtools}
\usepackage[colorlinks=true,citecolor=blue,linkcolor=blue]{hyperref}
\usepackage[caption=false]{subfig}
\usepackage{bbm}
\usepackage{comment}
\usepackage{hyperref}
\usepackage{upgreek}
\usepackage{esint}
\usepackage{soul}
\usepackage[normalem]{ ulem } 
\usepackage{floatrow}
\usepackage{verbatim}
\usepackage{braket}

\usepackage{array}
\newcommand{\PreserveBackslash}[1]{\let\temp=\\#1\let\\=\temp}
\newcolumntype{C}[1]{>{\PreserveBackslash\centering}p{#1}}

\newcommand{\rom}[1]{\uppercase\expandafter{\romannumeral #1\relax}}

\begin{document}

\title{Full programmable quantum computing with trapped-ions using semi-global fields}

\author{Yakov Solomons}\email{yakov.solomons@quantum-art.tech}
\author{Yotam Kadish}
\author{Lee Peleg}
\author{Jonathan Nemirovsky}
\author{Amit Ben Kish}
\author{Yotam Shapira}

\affiliation{Quantum Art, Ness Ziona 7403682, Israel}

\begin{abstract}
Trapped-ion quantum computing can utilize all motional modes of the ion-crystal, to entangle multiple qubits simultaneously, enabling universal computation with multi-qubit gates supplemented by single-qubit rotations. Using multiple tones to drive each ion individually induces Ising-type interactions, forming a multi-qubit gate, where the coupling matrix of all ion pairs is fully controllable. This reduces the total gate count while maintaining high fidelity, as opposed to traditional methods that rely on a single type of two-qubit gate, such as the well-known Mølmer–Sørensen gate. However, scaling to large ion chains, individual addressing can be technically challenging in terms of optical delivery and signal generation. We explore global and semi-global drives combined with single-qubit flips and show that these can reproduce the full set of multi-qubit gates. Although optimizing the combination of single-qubit flips is a computationally hard problem, we propose an efficient scheme to implement any desired couplings in large ion chains, yielding a concatenation scheme that uses at most $N/2$ multi-qubit gates, with $N$ being the number of ions. In addition, we show that by using $B<N$ independent semi-global fields, each driving a set of $N/B$ ions, the number of maximal multi-qubit gates is reduced to approximately $\frac{N^2}{B^2 (N-1)}$. We show how to design the driving fields that support these schemes and investigate their properties. Our results pave the way for efficient implementations of quantum algorithms in large-scale trapped-ion quantum systems.

\end{abstract}

\maketitle

\section{Introduction}
Universal quantum computing relies on the ability to execute any unitary operation with a finite sequence of gates. Typically, quantum computation is made universal by using arbitrary single-qubit operations combined with a unique type of two-qubit entangling gate, such as the Controlled-NOT gate or the well-known Mølmer–Sørensen (MS) gate ~\cite{molmer1999multiparticle,sorensen2000entanglement}, considered as the native gate in trapped-ion platforms. 

Trapped-ions systems have demonstrated high-fidelity quantum operations and good coherence properties~\cite{moses2023race,chen2024benchmarking,loschnauer2024scalable}, often relying on sequences of single- and two-qubit gates. However, these approaches typically underutilize the full potential of the system’s inherent all-to-all connectivity provided by the long-range Coulomb interaction. By harnessing this interaction along with the collective motional modes of the ion-crystal, trapped-ions systems can offer a more flexible and powerful approach by multi-qubit gates, enabling simultaneous entanglement of multiple qubit pairs within a single gate operation. This enables more efficient quantum processing by reducing the need to decompose complex unitaries into long two-qubit gate sequences. A larger gate set lowers the total gate count ~\cite{hoyer2003quantum,foxman2025random,bravyi2022constant,bassler2023synthesis,nemirovsky2025efficient}, preserving fidelity by minimizing errors from imperfections and decoherence, yielding more compact circuits that enhance speed and scalability.

For the implementation of a multi-qubit gate one effective approach involves addressing each ion with multiple frequency tones, thereby inducing Ising-type interactions that enable a pair-wise multi-qubit gate ~\cite{grzesiak2020efficient,lsf_paper}. This method offers full control over the coupling matrix governing interactions between all pairs of ions. However, scaling to larger ion chains poses significant practical challenges, as individual optical addressing becomes increasingly difficult ~\cite{chen2024low,niffenegger2020integrated,ParradoRodrguez2020CrosstalkSF,shen2013individual}; implementing an apparatus capable of independently modulating and delivering thousands of optical beams tightly focused to individual ions is technically demanding. Furthermore, arbitrary waveform signal generation per qubit drives the cost of constructing such a quantum computer and complicates its electronic classical control.

To address this issue we investigate an alternative strategy that leverages global and semi-global driving fields, that drive simultaneously few, $\sim10$, ions each. Combined with layers of single qubit $\pi$-flips we show that full programmability, similar to that available by individual ion addressing, is retained.
The proposal described below fits well to a quantum computer architecture based on long chains of trapped ions that are segmented by optical tweezers ~\cite{schwerdt2024scalable}. Furthermore we note that global driving fields not only mitigate the optical and signal generation challenges, but also relax constraints on ion-spacing. This in turn enables using larger ion-electrode distances in the ion trap, and acts to reduce motional heating rates. Similarly, constraints on thermal motion of the ions that is in the direction of the global beam's cross section can be relaxed as well.

The idea of adding layers of single-qubit $\pi$-flips was previously suggested in Refs.~\cite{hayes2014programmable,bassler2023synthesis,bassler2024general}, initially requiring $N(N-1)/2$ layers, and later improved to $N$ layers by utilizing all modes of motion ~\cite{Richerme_2025}. In this work, we further reduce the number of layers to $N/2$. Determining the optimal combination of single-qubit flips that minimizes the number of layers requires evaluating all $2^N$ possible configurations in each layer, a hard optimization problem that is computationally infeasible for large $N$ ~\cite{bassler2023synthesis}. To circumvent this challenge, we introduce an efficient scheme that uses pre-computation search which then enables the realization of arbitrary coupling matrices with a classical computational complexity of $\mathcal{O}(\mathrm{poly}(N))$ and a fixed worst-case number of layers. The key idea is to find configurations of single-qubit flips that generate a basis of matrices, allowing any desired coupling matrix to be decomposed into this basis. Notably, many such configurations form valid bases, and further optimization is performed to minimize the drive power. We demonstrate that this technique guarantees the total number of required multi-qubit gate layers does not exceed $\lceil N/2 \rceil$, significantly reducing the overhead associated with gate operations.

Furthermore, we explore a more generalized approach in which $B<N$ independent global fields are utilized to simultaneously drive sets of $N/B$ ions each (see Fig \ref{fig3}). With this we achieve an additional reduction in the maximum number of multi-qubit gates to approximately $\bigg\lceil N^2/(B^2 (N-1)) \bigg\rceil$.
Our study highlights the potential of global and semi-global driving schemes as a practical alternative to full individual addressing in large-scale trapped-ion quantum systems, paving the way for more efficient implementations of large scale quantum computers.

\section{Main results}
Considering first the MS gate, it drives $N$ ions by a global bichromatic field at $\omega_{\pm} = \omega_0 \pm (\nu + \xi)$, where $\omega_0$ is the ion qubit transition frequency, $\nu$ is the frequency of a selected motional mode, and $\xi$ is a small detuning from the sideband. The resulting effective interaction Hamiltonian reads ~\cite{shapira2020theory}
\begin{equation}
	H=\hbar \eta\Omega\cos(\omega t)  \bigg( {a}^\dagger e^{i \nu t} + {a} e^{-i \nu t} \bigg) 
	\sum_{n=1}^{N}O_n Z_{n},
	\label{H}
\end{equation}
where $\omega=\nu + \xi$, $a$ is the mode lowering operator, $\eta$  is the Lamb-Dicke parameter of the selected mode of motion, $\Omega$ is the Rabi frequency, $O_n$ is the participation of the n'th ion in the crystal, in the mode of motion, and $Z_n$ is the $Z$ Pauli operator of ion $n$. At the gate time $T=2\pi/\xi$, this Hamiltonian produces a $R_{zz}(-2\varphi)$ rotation in the spin subspace, with an idle operation on the phonons, with $\varphi\sim{\eta^2\Omega^2}/{\xi^2}$. E.g., assuming only two ions are illuminated, then at the gate time we have
\begin{equation}
	U_{MS}(\varphi)=\exp{\bigg(i\varphi Z_1Z_2\bigg)}.
\end{equation} 
This scheme has been generalized to multi-tone, multi-mode entangling gate ~\cite{webb2018resilient,shapira2020theory,shapira2018robust} in which the $N$ ions are driven with a global multichromatic laser field, containing $2M$ frequencies arranged in pairs $\omega_0\pm\omega_i$, that overlap the frequencies of the side-bands associated with all the normal-modes of motion of the ion-crystal. Each component has a phase $\phi_{\pm,i} = \pm \phi_i$, i.e., the average phase of each pair is $0$, and each pair has the same amplitude $\Omega r_i$, with $\Omega$ a characteristic Rabi frequency and $r_i \in \mathbb{R}$. This scheme results in an effective Hamiltonian of the form
\begin{equation}
	\begin{split}
		&H=\hbar\Omega\sum_{i=1}^{M}r_i\cos(\omega_i t + \phi_i)\\&
		\times\sum_{j=1}^{N}\eta_j\bigg({a}_j^\dagger e^{i \nu_j t} + {a}_j e^{-i \nu_j t} \bigg) 
		\sum_{n=1}^{N} O_n^{(j)} Z_{n},
		\label{H_global}
	\end{split}
\end{equation}
with $\eta_j$ being the Lamb-Dicke parameter of the j'th motional mode, $O_n^{(j)}$ being the participation of the n'th ion in the crystal, in the j'th mode of motion. With this generalized scheme one can implement any multi-qubit entangling gates of the form
\begin{equation}
U_g(\varphi)=\exp{\bigg(i\sum_j\varphi_j\sum_{n,m}M_{n,m}^{(j)}Z_nZ_m\bigg)}.
\label{U_g}
\end{equation}
Where $M_{n,m}^{(j)}=O_n^{(j)}O_m^{(j)}$ is a system-specific mode matrix. $\{\varphi_j\}$ is an arbitrary desired vector of $N$ phases which can be controlled by choosing the relative amplitudes of the different tones in the global driving field. While there are $N$ normal-modes of motion, the $M^{(j)}$ matrices in fact form a basis of dimension $N-1$ (due to the orthogonality of the modes), so only $N-1$ phases are independent. We remark that the formulation of Eq. \eqref{U_g} enables taking into account Intensity inhomogeneities of the driving beams, as shown in Ref. ~\cite{shapira2025programmable}

On the other hand, in individual addressing schemes, where each ion is driven independently, the attainable multi-qubit entangling gates have the form ~\cite{lsf_paper}
\begin{equation}
	U_s(\varphi)=\exp{\bigg(i\sum_{n,m}\varphi_{n,m}Z_nZ_m\bigg)},
\end{equation}
where $\varphi_{n,m}$ is an arbitrary desired $N\times N$ phase matrix, representing the coupling between all qubit pairs. Note that any coupling matrix $\varphi_{n,m}$ is part of the linear subspace of $N\times N$ real symmetric matrices, $\mathrm{Sym}(N)$ (due to the $n\leftrightarrow m$ symmetry). Furthermore, we assume the diagonal elements of $\varphi$ vanish as these correspond to trivial operations.

Comparing this with the global driving scheme, Eq. \eqref{U_g}, we see that while its vector phases $\varphi$, with its $N-1$ degrees of freedom (DoF), form a set of gates that is already much richer than the single Mølmer–Sørensen gate, it is still quadratically smaller than the set of gates formed by the linear subspace of $\varphi$ matrices in the individual addressing scheme. In other words, more DoF are needed to achieve arbitrary pairwise interactions.

\begin{figure}[t]
	\caption{Illustration of the main algorithm: layers of Ising-type multi-qubit entangling gates are sandwiched between single flips executed on different combinations of the $N$ qubits.}
	\includegraphics[width=1\textwidth]{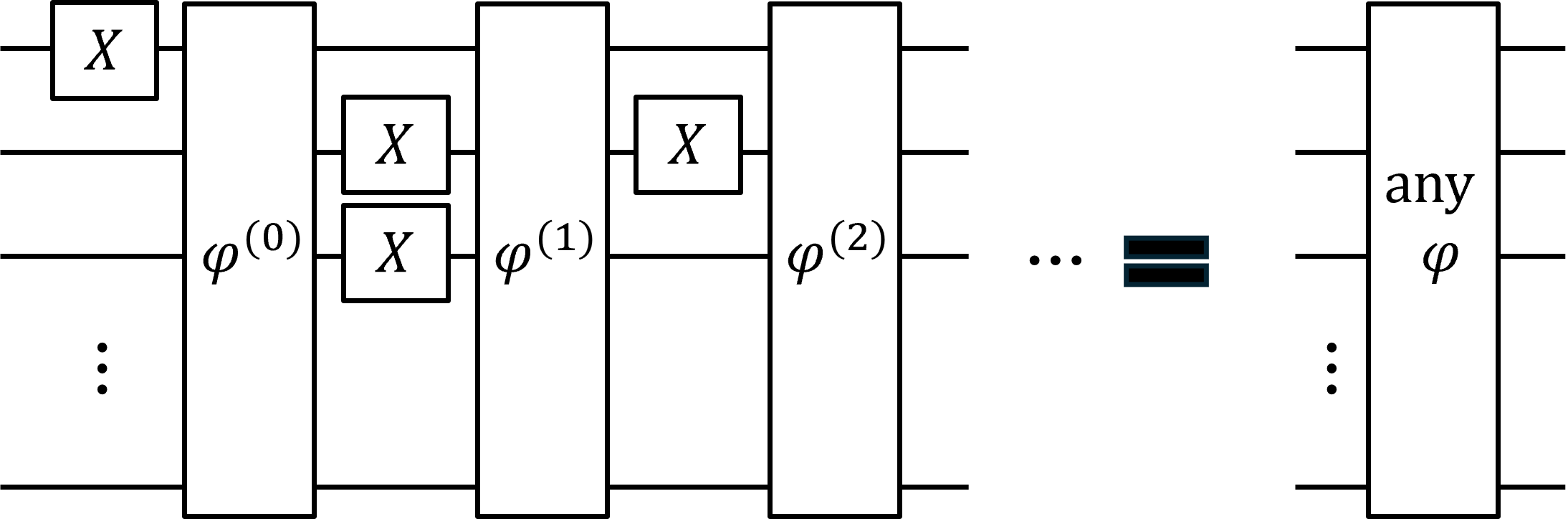}
	\label{fig1}
\end{figure}

In order to reproduce the full set of $\varphi_{n,m}$ phases with a global field, we interleave layers of global field gates and layers of single qubits flips (illustrated in Fig. \ref{fig1}). This makes use of individually addressing the ions, however only for simple single qubit $\pi$-rotations, which can be done much faster and easier than any entangling gate, with a myriad of methods e.g with steerable beams and with the assistance of light shifts, without stringent requirements of phase coherence between the gate and the single-qubit mechanism ~\cite{naegerl1999laser,lim2025design,sotirova2024low,flannery2024physical,manovitz2022trapped}. Thus we assume these rotations are less prone to errors and decoherence.

The single qubit layers are given by $X_{\mathbf{s}}=\Pi_n(X_n)^{s_n}$ with $s_n\in\{0,1\}$, such that $s_n$ determines whether the n'th ion is flipped or not. Adding these layers of single qubits flips enlarges the set of entangling gates that can be done with a global drive, as we will now show. The important fact about single qubit flips which are $\pi$-rotations around any axis in the $x-y$ plane, e.g, $R_x(\pi)$, is that they keep the system in the computational basis, therefore, an entangling gate sandwiched between $R_x(\pi)$ rotations only inverts the sign of the coupling phases.

\begin{equation}
	X_{\mathbf{s}}e^{\bigg(i\sum_{n,m}\varphi_{n,m}Z_nZ_m\bigg)}X_{\mathbf{s}}=e^{\bigg(i\sum_{n,m}\varphi_{n,m}S_{nm}Z_nZ_m\bigg)},
\end{equation}
with $S_{nm}=(-1)^{s_n}(-1)^{s_m}$. Similarly, adding another layer we have
\begin{equation}
	\begin{split}
		e^{\bigg(i\sum_{n,m}(\varphi_{n,m}^{(1)}S^{(1)}_{nm}+\varphi_{n,m}^{(2)}S^{(2)}_{nm})Z_nZ_m\bigg)}.
	\end{split}
\end{equation}
In general, with $L$ layers we have
\begin{equation}
	U=\Pi_{l}e^{\bigg(i\sum_{n,m}\varphi_{n,m}^{(l)}S_{n,m}^{(l)}Z_nZ_m\bigg)}=e^{\bigg(i\sum_{n,m}\Phi_{n,m}Z_nZ_m\bigg)}, 
\end{equation}
with
\begin{equation}
	\Phi_{n,m}=\sum_{l}\varphi_{n,m}^{(l)}S_{n,m}^{(l)}.
\end{equation}
Decomposing $\varphi_{n,m}^{(l)}$ with the given basis of the mode structure, the set of matrices $M^{(j)}$, we can write $\varphi_{n,m}^{(l)}=\sum_j\alpha_j^{(l)}M_{n,m}^{(j)}$, so
\begin{equation}
	\Phi_{n,m}=\sum_{l}\sum_j\alpha_j^{(l)}M_{n,m}^{(j)}S_{n,m}^{(l)},
	\label{phi}
\end{equation}
with $\alpha_j^{(l)}$ the entanglement phase accumulated by the j'th mode of motion at the l'th layer. While the set of matrices $M^{(j)}$ are only a small subspace of the linear space $\mathrm{Sym}(N)$, the matrices $\tilde{M}^{(j,l)}\equiv M^{(j)}S^{(l)}$ are a larger set, thus, one can hope to construct a  desired target coupling matrix $\varphi_{n,m}^{\mathrm{target}}$ by choosing the correct combination of the single qubit flips $X_{\mathbf{s}}$ and the coefficients $\alpha_j^{(l)}$, such that $\varphi_{n,m}^{\mathrm{target}}\approx\Phi_{n,m}$. However, determining the optimal set of single-qubit flips naively requires searching through all $2^N$ possible combinations for each layer separately ($2^{NL}$ in total), a computationally infeasible optimization problem per $\varphi^{\mathrm{target}}$ ~\cite{bassler2023synthesis}. 

To address this issue, we propose a new approach where the search needs to be performed only once in advance. The idea is that one should look for a combination of single qubit flips such that the generated basis of matrices $\{\tilde{M}^{(j,l)}\}$ will cover the full linear space $\mathrm{Sym}(N)$. Since there are $N(N-1)/2$ basis elements of $\mathrm{Sym}(N)$ (neglecting the physical irrelevant diagonal terms), and each layer of multi-qubit gates covers a $N-1$-dimensional subspace of $\mathrm{Sym}(N)$, from a simple count of DoF we expect that $\lceil N/2\rceil$ layers of multi-qubit gates will suffice.

We do so by writing each mode matrix $M^{(j)}$ as a $N(N-1)/2$ vector, collecting the set of $N$ vectors as columns of a matrix $A$, and then looking for a combination of single qubit flips $X_{\mathbf{s}}$ such that adding the transformed vectors $M^{(j)}S^{(l)}$ to our current set yields the largest possible linearly independent basis, i.e., maximizing the rank of the collection matrix $A$. For relatively small qubit numbers, one may employ an exhaustive search over all possible combinations of single-qubit flips in each layer. For larger system sizes, one can adopt an optimization procedure in which $1$ or $2$ single-qubit flips are added at each step, with the final configuration chosen so as to maximize the cost function, defined as the rank of the matrix $A$. This process continues until a full basis of $N(N-1)/2$ vectors is constructed, spanning the full linear space $\mathrm{Sym}(N)$, ensuring that any arbitrary coupling matrix is covered. Once a set of $\lceil N/2\rceil$ combinations of single qubit flips has been determined, the only cost per gate is a projection of $\varphi_{n,m}^{\mathrm{target}}$ on this basis to determine the coefficients, $\alpha_j^{(l)}$ per gate. This boils down to finding the pseudo-inverse of $A$, which scales as $\mathcal{O}(\mathrm{poly}(N))$.

\begin{figure}[t]
	\caption{Number of multi-qubit gate layers (points) required to implement arbitrary multi-qubit gates as a function of the number of qubits, for various choices of semi-global fields $B$. Our DoF-based estimation is shown as well (dashed).}
	\includegraphics[width=1\textwidth]{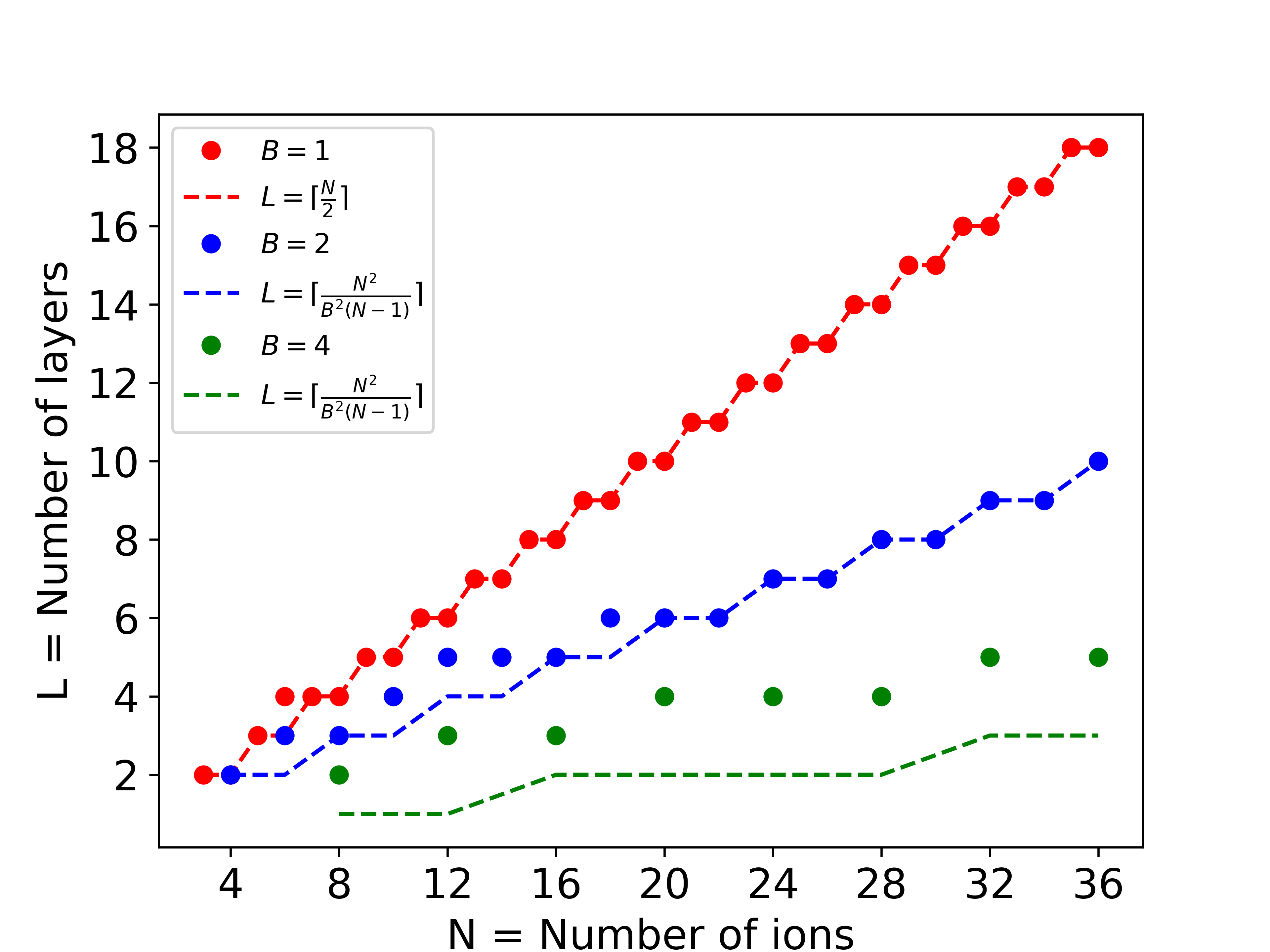}
	\label{fig2}
\end{figure}

In practice, many sets of basis for $\mathrm{Sym}(N)$ may be found. As the drive power is limited, and gate infidelity often increases with drive power, a practical choice then is to prioritize a basis via minimization of the $L_\infty$ norm of $\mathrm{pinv}(A)$ ($\mathrm{pinv}:=$pseudo inverse). This optimization, which selects maximally separated vectors, ensures small $\alpha_j^{(l)}$ coefficients which translates to minimum drive power. Indeed, solutions involving up to $\lceil N/2\rceil$ layers of multi-qubit gates are almost always found, as shown in Fig. \ref{fig2} ($B=1$ red dataset), where we plot the number of layers as a function of the number of qubits, calculated with our presented method (points), and compared to our DoF-based estimation $\lceil N/2\rceil$ (dashed line). We remark that $\lceil N/2\rceil$ is an upper bound on the number of multi-qubit layers, and that further optimizations may be employed on a gate-by-gate basis to reduce the number of layers further. 

\begin{figure}[t]
	\caption{(a) Illustration of a line of ions illuminated by two semi-global beams. (b) The corresponding coupling matrix $\Phi$ is split to $B=2$ global sub-beams, which generate intra- (red and blue) and inter- (mixed red and blue) sub-beam couplings. The lower triangular part is shaded as it is irrelevant.}
	\includegraphics[width=1\textwidth]{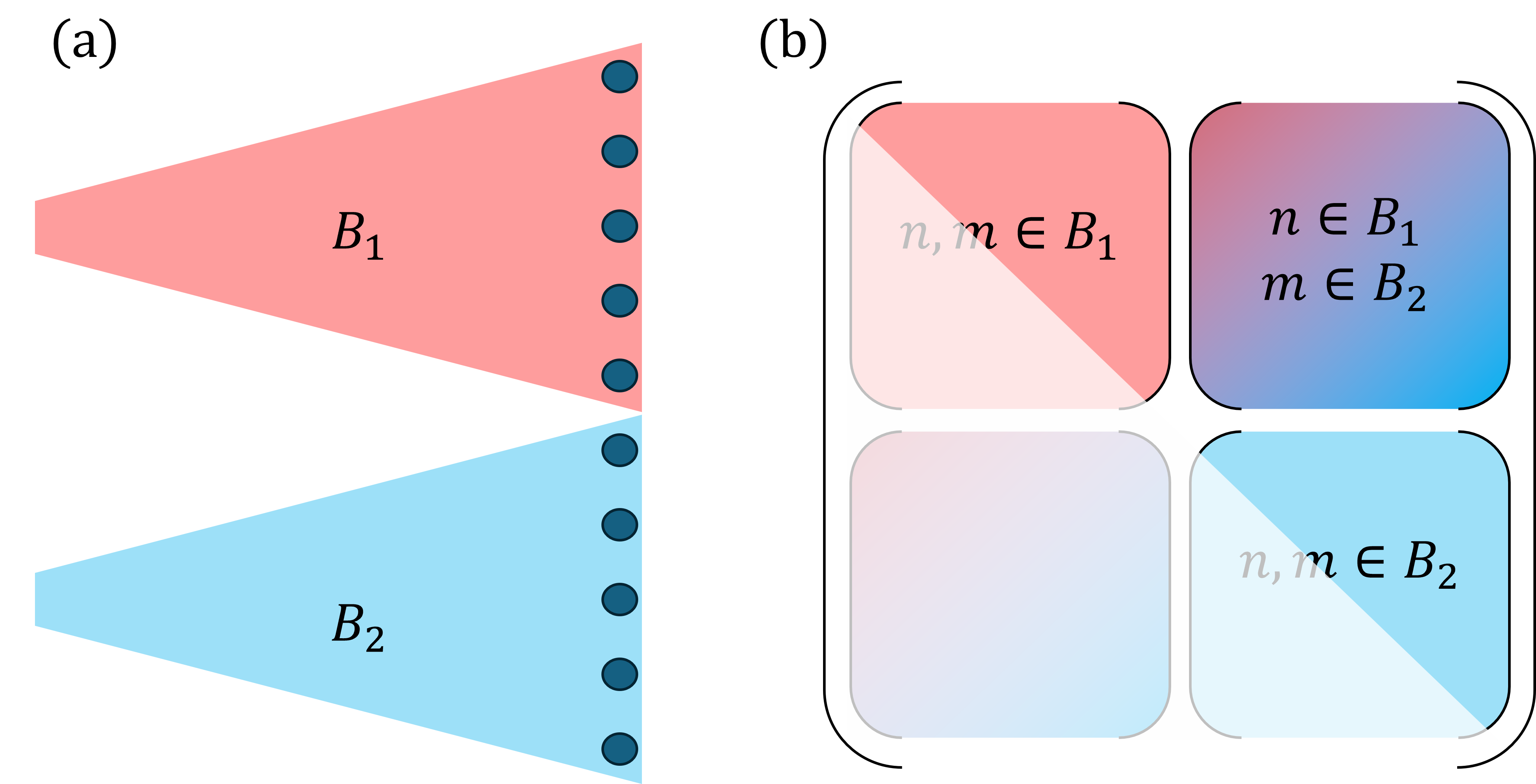}
	\label{fig3}
\end{figure}

Nevertheless, $\lceil N/2\rceil$ can still be considerable for large values of $N$, therefore an additional generic reduction of the number of multi-qubit layers is required. We perform this further reduction by splitting the global beam to $B$ sub-global beams, each of which homogeneously illuminates a section of $N/B$ ions. 
Fig. \ref{fig3}b shows an example of a coupling matrix $\Phi_{n,m}$, that is split using $B=2$ sub-beams. The upper triangle (red) is the coupling between pairs of qubits illuminated by the $B_1$ beam, the bottom triangle (blue) is the coupling between pairs of qubits illuminated by the second beam $B_2$, and the square (color gradient) is the coupling between qubits in different segments. These parts correspond to the following equations (compare to Eq. \eqref{phi})

\begin{equation}
	\label{phi_split}
\begin{split}
&\Phi_{n,m\in B_1}=\sum_{l}\sum_j\alpha_j^{l}M_{n,m\in B_1}^{(l,j)}S_{n,m\in B_1}^{(l,j)}
\\&\Phi_{n,m\in B_2}=\sum_{l}\sum_j\beta_j^{l}M_{n,m\in B_2}^{(l,j)}S_{n,m\in B_2}^{(l,j)}
\\&\Phi_{\substack{n\in B_1\\m\in B_2}}=\sum_{l}\sum_j\gamma_j^{l}M_{\substack{n\in B_1\\m\in B_2}}^{(l,j)}S_{\substack{n\in B_1\\m\in B_2}}^{(l,j)}.
\end{split}
\end{equation}
We note that, in practice the coefficients $\alpha,\beta,\gamma$, are not independent, since $\gamma$ represents the coupling phases between qubits in different segments and is determined by the same two beams that also generate the intra-segment phases $\alpha$ and $\beta$. Nevertheless, we can treat them as freely chosen independent parameters, as it is always possible to find a suitable set of tones for each beam that satisfies their underlying relation, as we explain below.

This approach is generalized to arbitrary choice of $B$ beams, resulting always in two types of couplings – intra sub-beam coupling, i.e pairs of qubits driven by the same semi-global beam, of which there are $B\frac{N}{B}(\frac{N}{B}-1)/2$ phases ($B$ triangular sections, each with ${\frac{N}{B}(\frac{N}{B}-1 )}/{2}$ elements), and inter sub-beam coupling, i.e pairs of qubits driven by different semi-global beams, of which there are ${B(B-1)}/{2}(\frac{N}{B})^2$ phases (${B(B-1)}/{2}$ square sections, each with $(\frac{N}{B})^2$ elements). 

We consider the set of phases between two sub-beams (which is larger than the number of intra-beam couplings) and invoke a similar count of DoF as above. This yields a lower bound on the number of multi-qubit layers, namely, $\lceil \frac{N^2}{B^2(N-1)}\rceil$ (note that this bound does not converge to the bound $\lceil N/2\rceil$ obtained for a single global beam $B=1$).

Using the same type algorithm, we are indeed able to reduce the number of multi-qubit layers, as shown in Fig. \ref{fig2} (blue and green plots). For example, a chain of $N=36$ ions and $B=4$ sub-beams requires at most only $5$ applications of multi-qubit gates, per $\varphi_{n,m}^{\mathrm{target}}$. Note that as the number of beams increases, the lower bound $\lceil \frac{N^2}{B^2(N-1)}\rceil$ is not always achieved. This occurs because, when the coupling matrix is divided into smaller parts, single-qubit flips alone may not generate enough independent vectors to span the entire linear space. Nevertheless, this method ensures that the number of layers will be as low as possible for the construction of an arbitrary coupling matrix. In addition, the basis construction method is only an upper bound, and further optimizations may be employed on a gate-by-gate basis to reduce the number of layers further, as noted above. 

\section{Drive properties for semi-global fields}

To investigate the impact of semi-global fields on the implementation of general entangling gates, we adopt a method to compute the drive tones for arbitrary gates and addressing patterns. The procedure is compatible with any pulse parameterization; here we follow the formulation of Eq. \eqref{H_global}, wherein  each beam is specified by its set of amplitudes $\vec{r}^{b}$. We then optimize the drive parameters for configurations with $B\in\{2,\ldots,N\}$. In particular, we extend the Large-Scale Fast (LSF) method ~\cite{lsf_paper} from fully individual driving ($B=N$) to semi-global driving ($B<N$) to obtain gate solutions for various addressing patterns. 

In the semi-global driving scheme, the spectral components of different beams are no longer fully independent due to inter-beam coupling between ions in different segments. These couplings introduce additional constraints on the control signal. Nevertheless, our extensions to the LSF method can efficiently find solutions sufficient for high-fidelity gate implementation.

To incorporate our multi-layer gate construction within the LSF framework, we proceed as follows: Once a sequence of single-qubit flips is specified we use Eq. \eqref{phi_split} (or its generalized version in case of $B>2$) to decompose a given target gate to $L$ layers. Each layer, sandwiched between single-qubit operations, corresponds to an intermediate target gate of the form $\varphi_{n,m}^{l}=\sum_j\alpha_j^{l}M_{n,m}^{(l,j)}S_{n,m}^{(l,j)}$. Then, the physical control signal \( r^{(l)} \) responsible for generating the target-layer unitary \( U = \exp\left( \sum_{n,m} \varphi_{n,m}^{(l)} Z_n Z_m \right) \) is constructed by the LSF optimization procedure. Specifically, the control modulation for beam \( b \) is represented in a Fourier basis
\begin{equation}
	w_b(t) = \sum_{m=1}^M r_m^{(b)} \cos(\omega_m t + \phi_m^{(b)}),
\end{equation}
where \( \{\omega_m\}_{m=1}^M \) is a uniform frequency grid spanning the motional mode spectrum of the ion crystal, and \( \vec{r}^{(b)} \), \( \vec{\phi}^{(b)} \) are the amplitudes and phases of the frequency components for beam \( b \).

The gate synthesis problem is formulated as a quadratically constrained minimization:
\begin{equation}
	\hat{R}^{(l)} = \arg\min_R \|R\| \quad \text{subject to} \quad R_n A_{nm} R_m = \varphi_{n,m}^{(l)},
	\label{Eq13}
\end{equation}
where $A_{nm}$ is the bilinear kernel coupling the drive spectra of ion $n$ and $m$ to the resulting entangling phase, and \( R_n \) represents the drive on ion \( n \), expressed in coordinates lying in the kernel of the displacement constraints. Although Eq. \eqref{Eq13} forms a NP-hard optimization problem, LSF efficiently finds approximate solutions in polynomial time, sufficient for high-fidelity implementation. 

For semi-global fields only \(B<N\) drives are individually determined. The quadratic constraints in Eq. \eqref{Eq13} above can be adapted by introducing a labeling map \(O_{b,n}\), such that $O_{b,n}=1$ if qubit $n$ is driven by beam $b$ and $0$ otherwise, yielding the set of constraints \(R_{b_1}O_{b_1,n}A_{nm} O_{b_2,m} R_{b_2}=\varphi_{n,m}^{l}\) for every qubit-pair. This provides a natural generalization of LSF optimization to semi-global beams. While there is strong evidence for the feasibility of satisfying the constraints in Eq. \eqref{Eq13}, as in practice LSF always provides solution to the gate-design problem, we note that in the adiabatic gate regime one can satisfy these constraints by-hand (see Appendix \ref{App}).

\begin{figure}[t]
	\caption{Average ratio of total Rabi frequency per layer $|r|$ to the nuclear norm $\Omega_{\mathrm{nuc}}$ of the target gate coupling matrix, plotted as a function of the number of driving beams $B/N$, for $N=12$ (red), and $N=36$ (blue). The increase in required power for fewer beams reflects the added constraints from inter-beam coupling in the semi-global scheme.}
	\includegraphics[width=1\textwidth]{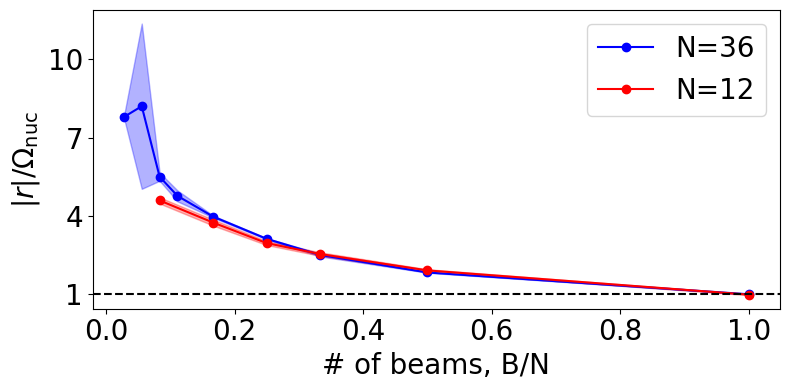}
	\label{fig4}
\end{figure}

It is instructive to compare the performance of the proposed semi-global scheme to that of individual addressing in terms of the power required to implement a desired entangling gate. Ref. \cite{lsf_paper} conjectured an estimate for the optical power required to drive arbitrary entangling maps between ions in a single crystal, scaling as the $L_1$ norm of the entangling map $\varphi$ eigenvalues. This estimate is known as the nuclear norm, $\Omega_{\mathrm{nuc}}$. In case of semi-global beams one can expect that additional power will be needed due to the additional constraints that take care of the inter-beam coupling. Indeed, we find that the total Rabi frequency $|r|$ per layer is increased as the number of driving beams decreases. This is shown in Fig. \ref{fig4} with the ratio $|r|/\Omega_{\mathrm{nuc}}$ of random gates (averaged over 10 different gates and over all layers) plotted as a function of the number of beams, with $\Omega_{\mathrm{nuc}}$ being the nuclear norm of the coupling matrix per layer. Importantly, this increase in power is not prohibitive—it represents a modest and reasonable trade-off for the substantial experimental simplification gained by avoiding full individual optical addressing.

\section{Conclusions}
We have introduced a scheme for implementing universal quantum computation in large-scale trapped-ion systems using semi-global driving fields combined with layers of single-qubit operations. This approach avoids the need for full individual optical addressing while still enabling the synthesis of arbitrary multi-qubit entangling gates. By efficiently generating a basis of coupling matrices through carefully selected single-qubit flip configurations, we showed that any desired target gate can be constructed with at most $\lceil{N/2}\rceil$ layers of multi-qubit gates. To further reduce the circuit depth, we extended the method to configurations with $B<N$ independent driving beams, each illuminating a subset of the ion chain. This generalization enables a systematic tradeoff between hardware complexity and gate-layer count, achieving a reduction in the number of multi-qubit gates to approximately $\bigg\lceil\frac{N^2}{B^2 (N-1)}\bigg\rceil$. We also adapted the LSF method for the semi-global regime, allowing efficient computation of the required control signal. Our results present a practical path toward implementing efficient and high-fidelity quantum algorithms in large-scale trapped-ion systems without requiring full individual control over each qubit.

\begin{acknowledgments}
We thank Roee Ozeri for valuable conversations and insights.
\end{acknowledgments}

\bibliography{references}

\cleardoublepage
\onecolumngrid

\appendix

\section{An analytical expression for the adiabatic regime}
\label{App}
In the main text, we noted that it is not obvious that the LSF method can always find a solution to the gate-design problem subject to the constraints in Eq. \eqref{Eq13}. This difficulty arises because, in the semi-global driving scheme, the spectral components of different beams are no longer fully independent, due to inter-beam couplings between qubits in different segments. In particular, each sub-beam must simultaneously generate all intra-beam entangling phases while also contributing to the inter-beam phases. Although in practice the LSF reliably finds a solution, in the adiabatic gate regime the constraints can in fact be satisfied analytically. To demonstrate this, we consider a gate operated in the extreme adiabatic limit, i.e., at very long gate durations $\tau$.

For example, in this regime drive tones that are symmetrically detuned around the red and blue sideband transition of an all-to-all mode of motion and are detuned by a frequency $\xi_n=n\xi_0=2\pi n/\tau$ generate the entanglement phase, $\varphi_{j,k}=\varphi_0r_{j,n}r_{k,n}/n$, with $r_{j,n}$ the amplitude of the tone $n\xi_0$ in the j'th sub-beam and $\varphi_0$ a predetermined phase. Furthermore, in this regime different tones do not interfere, i.e. the entanglement phase does not have a contribution that scales as $r_{j,n}r_{j',m}$ (mixing between different tones). 

Now, as a simple example we consider the case $B=2$ with a single intra-beam entanglement phase per beam, $\varphi_a\varphi_0$ and $\varphi_b\varphi_0$, and a single inter-beam entanglement phase, $\varphi_{a,b}\varphi_0$. 
We show a solution in the adiabatic regime with three tones in beam (a): $\xi_{n_1}$, $\xi_{n_2}$ and $\xi_{n_3}$, and three tones in beam (b): $\xi_{n_2}$, $\xi_{n_3}$ and $\xi_{n_4}$. This scenario requires the solution to satisfy the following 3 constraints
\begin{equation}
	\begin{split}
		&\frac{r_{a,1}^2}{n_1} +\frac{r_{a,2}^2}{n_2} + \frac{r_{a,3}^2}{n_3}=\varphi_a,\\
		&\frac{r_{b,2}^2}{n_2} +\frac{r_{b,3}^2}{n_3} + \frac{r_{b,4}^2}{n_4}=\varphi_b,\\
		&\frac{r_{a,2}r_{b,2}}{n_2} +\frac{r_{a,3}r_{b,3}}{n_3}=\varphi_{a,b}.
	\end{split}
\end{equation}
To solve it, we set $\frac{r_{a,1}^2}{n_1}=\varphi_a$ and $\frac{r_{b,4}^2}{n_4}=\varphi_b$ (note that the sign of $n_1$ and $n_4$ accommodates for the sign of the $\varphi$'s), and furthermore we set $\frac{r_{a,2}^2}{n_2}=-\frac{r_{a,3}^2}{n_3}$ and $\frac{r_{b,2}^2}{n_2}=-\frac{r_{b,3}^2}{n_3}$. For example, for $n_2>0$ and $n_3<0$, these constraints are satisfied with $r_{a,2}=\sqrt{|n_2|/|n_3|}r_{a,3}$ and $r_{b,2}=-\sqrt{|n_2|/|n_3|}r_{b,3}$, such that the third constraint reads, $2\frac{r_{a,3}r_{b,3}}{n_3}=\varphi_{a,b}$, which is easily satisfiable as both $r_{a,3}$ and $r_{b,3}$ are unconstrained.

\end{document}